\documentclass{aastex6}

\lefthead{Wada}
\righthead{Multi-phase gas around AGN}

\newcommand{\bea}{\begin{eqnarray} }
\newcommand{\eea}{\end{eqnarray}}

\begin{document}

\title{ Multi-phase Nature of a Radiation-Driven Fountain with Nuclear Starburst in a Low-mass Active Galactic Nucleus  }
 
\author{%
Keiichi \textsc{Wada}\altaffilmark{1}
}%
\affil{Kagoshima University, Kagoshima 890-0065, Japan}
\email{wada@astrophysics.jp}

\author{Marc \textsc{Schartmann}}%
\affil{Centre for Astrophysics and Supercomputing, Swinburne University of Technology, P.O. Box 218, Hawthorn, Victoria 3122, Australia}

\author{Rowin \textsc{Meijerink}}
\affil{	Leiden Observatory, P.O. Box 9513
NL-2300 RA  Leiden
The Netherlands
}

 \altaffiltext{1}{Research Center for Space and Cosmic Evolution, Ehime University, Japan}

\begin{abstract}
The structures and dynamics of 
molecular, atomic, and ionized gases are studied around a low-luminosity active galactic nucleus (AGN) 
with a small ($2\times 10^6 M_\odot$) black hole 
using three-dimensional (3D) radiation hydrodynamic simulations.
We studied, for the first time, the non-equilibrium chemistry for the X-ray dominated region
in the ``radiation-driven fountain" \citep{wada2012} with supernova feedback.
A double hollow cone structure is naturally formed without postulating
a thick ``torus'' around a central source.
The cone is occupied with an inhomogeneous, diffuse ionized gas and surrounded
by a geometrically thick ($h/r \gtrsim1$) atomic gas.
Dense molecular gases are distributed near
the equatorial plane, and energy feedback from supernovae  enhances their
scale height.
Molecular hydrogen exists in a hot phase 
($ >$ 1000 K)  as well as in a cold ($ < 100$ K), dense ($ >10^3$ cm$^{-3}$) 
phase. The velocity dispersion of H$_2$ in the vertical direction is 
comparable to the rotational velocity, which is consistent with 
near infrared observations of nearby Seyfert galaxies.
Using 3D radiation transfer calculations for the dust emission, we find
polar emission in the mid-infrared band (12 $\mu {\rm m}$), which is associated with
bipolar outflows, as suggested in recent interferometric observations of nearby AGNs.
If the viewing angle for the nucleus is larger than 75$^\circ$, the spectral energy distribution
is consistent with that of the Circinus
galaxy. The multi-phase interstellar medium observed
in optical/infrared and X-ray observations  is also discussed.

\end{abstract}


\section{INTRODUCTION}
Recent observations of nearby active galactic nuclei (AGNs) suggest that  
the long-standing standard picture, i.e.,  bright emission from a
central source is obscured by a dusty ``donut-like'' torus,  needs to be reconsidered.
For example, mid-infrared (MIR) interferometric observations showed that 
 the bulk of MIR  is emitted from the dust in the polar region, not 
from the dusty torus \citep{tristram2014, asmus2016}.
However, it is not clear how this MIR emission  is related with  the double (or single) 
hollow cones of ionized gas that are often observed in many
AGNs. The observed properties of AGNs are related to 
the conditions of the interstellar  medium (ISM) near the nucleus, but
its multi-phase structure in the context of the standard picture is still unclear.

Recently, we have proposed a novel mechanism of the obscuring structures around 
AGNs in which outflowing and inflowing gases are driven by 
radiation from the accretion disk, forming 
a geometrically thick disk on the scale of a few to tens of parsecs \citep{wada2012} (hereafter W12).
The quasi-steady circulation of gas, i.e., the ``radiation-driven fountain", may
 obscure the central source, and thereby
  the differences in the spectral energy distributions (SEDs) 
 of typical type-1 and type-2 Seyfert galaxies
 are consistently explained \citep{schartmann2014} (hereafter Sch14).  
We also showed in \citet{wada2015} (hereafter W15) that the observed properties of obscured AGNs
changes as a function of their luminosity due to the fountain flows, which is consistent with recent observations
\citep[e.g.,][]{burlon2011, lusso2013, ueda2014, buchner2015, aird2015}. 

There are many instances in which AGNs are associated with
circumnuclear starbursts \citep[e.g.,][]{davies2007, hicks09, chen09, imanishi2011, Diamond2012, durre2014, Woo2012, esquej2014, bernhard2016}. 
{The nuclear starbursts themselves could inflate the circumnuclear disk and 
obscure the central source \citep{wada02, wada09}.
Therefore, the interplay between the radiation-driven fountain and the feedback from nuclear starbursts 
should be an important process to be understood for the multi-phase nature of the ISM around AGNs.


{{ In this Letter, we apply the radiation-driven fountain model to the case of 
a low black hole (BH) mass  ($\sim 10^6 M_\odot$), which was not studied in our previous papers.
We expect that the nuclear starburst would have more impact on structures of the circumnuclear 
disk around smaller BHs \citep{wada02, wada2004}.
For the first time, we solve  the X-ray dominated region (XDR) chemistry with
three-dimensional (3D) gas dynamics under the effect of the radiation from AGN and 
and the energy feedback from supernovae (SNe).
The selected parameters are close to those in the nearest Seyfert galaxy, the Circinus galaxy, 
and we compare the results with multi-wavelength observations using 3D radiative transfer calculations. 

 }

\section{NUMERICAL METHODS AND MODELS}
\subsection{Numerical methods}
We follow the same numerical methods as those given in W12 and W15, i.e.
3-D  Eulerian hydrodynamic code 
 with a uniform grid that accounts for radiative feedback processes from the central source using a ray-tracing method.
One major difference from our previous papers is that we include the non-equilibrium XDR chemistry \citep{maloney96,
meijerink05} for 256$^3$ zones (resolution of 0.125 pc). 
In the present models, self-gravity of the gas is ignored because it is not essential for the gas dynamics in the radiation-driven fountain (see also \citet{namekata2016}).
The supenova feedback is implemented based on \citet{wada01}, in which the energy from a supernova is 
injected in a randomly selected grid cell near the equatorial plane. 

We assumed that the radiation flux caused by the accretion disk 
is non-spherical (see below), and the flux 
is calculated for all grid cells using $256^{3}$ rays. 
The outflows are driven mainly by the radiation pressure of dust and X-ray heating. 
A cooling function for 20 K $< T_{gas} < 10^{8}$ K \citep{meijerink05, wada09} and 
solar metallicity are assumed.
We include photoelectric heating due to a uniform far ultraviolet (FUV)  field with $G_{0} = 1000$, where $G_0$ is the incident FUV field normalized to the local interstellar value. 

If temperature of the dust irradiated by a central source $T_{\rm dust}$ exceeds $1500\,{\rm K}$, we assume that the dust is sublimated, 
and no dust is assumed for $T_{gas} > 10^5 $ K due to dust sputtering. 
Here, we assume that the ISM in the central several tens of parsecs is optically thin 
with respect to the re-emission from the hot dust.
The SED of the AGN and the dust absorption cross-sections are taken from \citet{laor1993}.
{{ In the hydrodynamic calculations, the dust is not treated as an independent component, but
assumed to move together with the gas.}
We use a standard galactic dust model \citep{schartmann2011} with a dust-to-gas ratio of  0.01.

We used a selection of reactions from the chemical network described by \citet{meijerink05, adamkovics2011}
for 25 species:  H, H$_2$, H$^+$,  H$_2^+$, H$_3^+$, H$^-$, e$^-$, O, O$_2$, O$^+$,  O$_2^+$, O$_2$H$^+$, OH,
OH$^+$, H$_2$O, H$_2$O$^+$, H$_3$O$^+$, C, C$^+$, CO, Na, Na$^+$, He, He$^+$, and HCO$^+$.
At every time step, the gas density, gas and dust temperatures, and ionization parameters in the 256$^3$ grid cells
are passed to the chemistry module. The chemistry module returns the abundances of the species,
and they are advected based on the gas velocity \footnote{To save computational memory, we follow the advection only for
H, H$^+$, H$_2$, O, O$^+$, H$_2$O, OH, C, C$^+$, and $e^-$}. 

\subsection{Initial Conditions and Model Set-up}
{{ In contrast to our previous papers (W12, W15, and Sch14),
we assume a smaller black hole mass ($2\times 10^6 M_\odot$) in this study,  and
other parameters are selected to fit with the nearest AGN, the Circinus galaxy.
Then the numerical results can be directly compared with observations  (see \S 4). }
We follow the evolution of a rotating gas disk (total gas mass
is $ 2 \times 10^6 M_\odot$) in a
fixed spherical gravitational potential under the influence of radiation from 
the central source (the Eddington ratio of 0.2 and the bolometric luminosity of $L_{bol}=5 \times 10^{43}$ erg s$^{-1}$ 
are fixed during the calculation). 
 We assume a time-independent external potential
$\Phi_{\rm ext}(r) \equiv -(27/4)^{1/2}[v_1^2/(r^2+a_1^2)^{1/2}+v_2^2/(r^2+ a_2^2)^{1/2}]$, where $a_1 = 100$ pc, $a_2 =
2.5$ kpc, $v_1 = 147$ km s$^{-1}$, and $v_2 = 147$ km s$^{-1}$.
To prepare quasi-steady initial conditions without
radiative feedback, we first evolve axisymmetric and rotationally supported thin
disks with uniform density profiles.
After the thin disks are dynamically settled, the radiation 
feedback and SNe feedback ({{ the supernova rate is a free parameter 
and is fixed during the calculation}) are turned on.

The ultraviolet flux is assumed to be $F_{UV}(\theta) \propto \cos \theta (1+2\cos \theta)$ \citep{netzer1987},
where $\theta$ denotes the angle from the rotational axis ($z$-axis).
The X-ray radiation, on the other hand, is assumed to be spherically symmetric  \citep{netzer1987, xu2015}.
The UV and X-ray fluxes are calculated from the bolometric luminosity \citep{marconi2004}. 
The total X-ray luminosity (2--10 keV) is $L_X = 2.8\times 10^{42} $ erg s$^{-1}$.

Although the calculations are  fully three-dimensional (i.e., without assuming any symmetries), 
the direction of the emerging non-spherical radiation
is assumed to be parallel to the rotational axis of the circumnuclear gas disk in the present model
(see W15 for non-parallel cases).

\section{RESULTS}
Figure 1 shows the distributions of three phases, i.e., atomic, molecular and ionized hydrogen, as well as
the gas temperature in a quasi-steady state.
The distribution of neutral hydrogen (H$^0$) 
shows a hollow cone structure surrounded by a geometrically thick disk,
similar to that typically observed in the radiation-driven fountain models (W12, W15). 
The bipolar outflow of the low density gas is directly caused by the
radiation pressure and X-ray heating due to the AGN radiation.

The high-density molecular gas forms a clumpy disk similar to that in the previous starburst model \citep{wada02, wada09}.
A clumpy disk of H$_2$ is embedded in the geometrically thick disk of H$^0$. 
Here, H$^+$  is mostly distributed outside the thick neutral disk and 
in the bipolar outflows, which are deficient in H$^0$ and H$_2$.  
The maximum velocity of the outflow is  $\sim 500$ km s$^{-1}$ near the center, 
but it slows down to $\sim 100$ km s$^{-1}$ in the outer region.
 In the majority of the high density thin disk, 
the temperature of the gas is less than 100 K, except for the hot regions caused by supernovae.

\begin{figure}[h]
\centering
\includegraphics[width = 12cm]{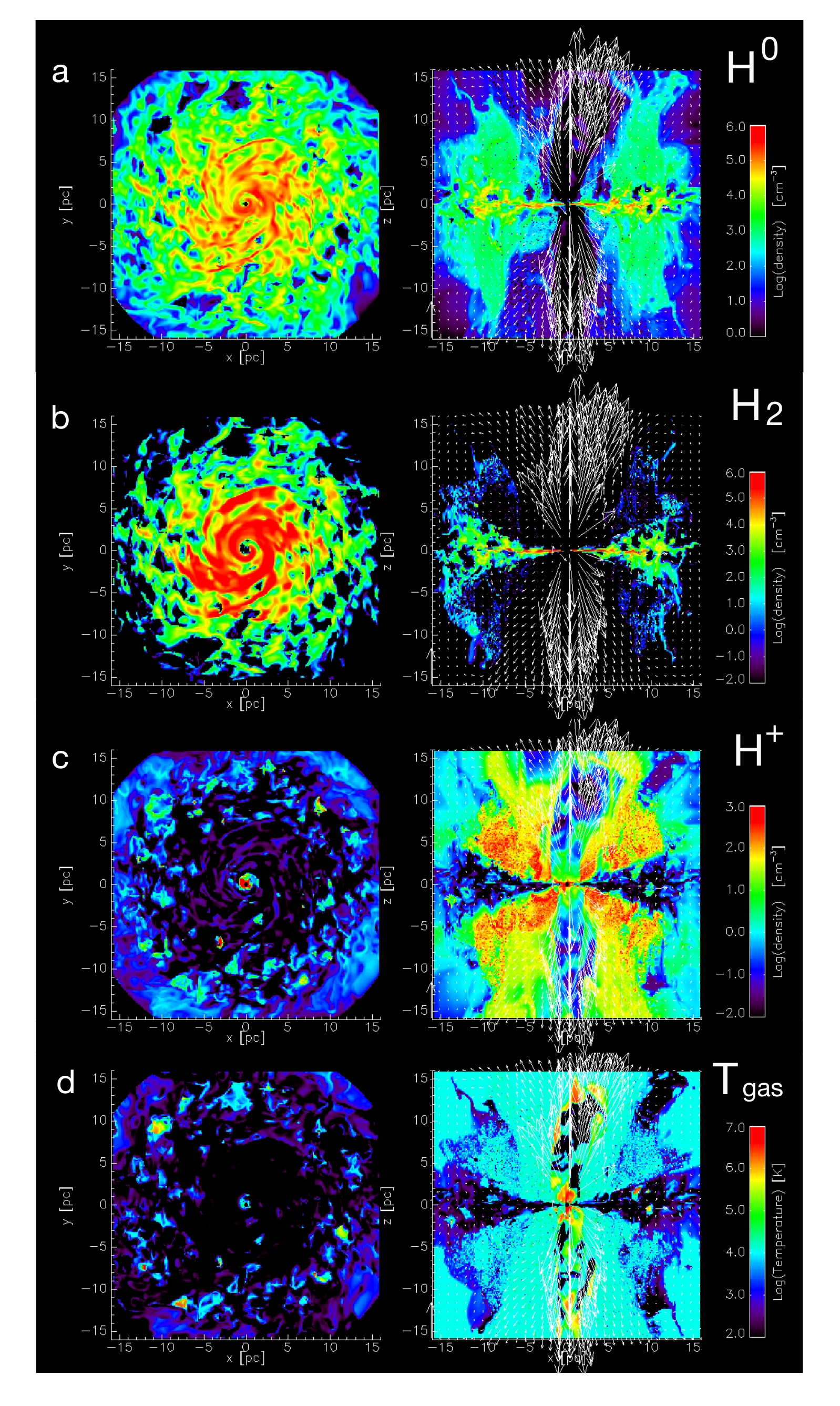} 
\caption{Distributions of (a) H$^0$,  (b) H$_2$,   (c) H$^+$, and (d) gas temperature on
the equatorial plane and $x-z$ plane. The arrows represent velocity vectors of the gas. The average supernova rate is 0.014 yr$^{-1}$.}
\label{wada_fig: phase}
\end{figure}

\begin{figure}[h]
\centering
\includegraphics[width =  10cm]{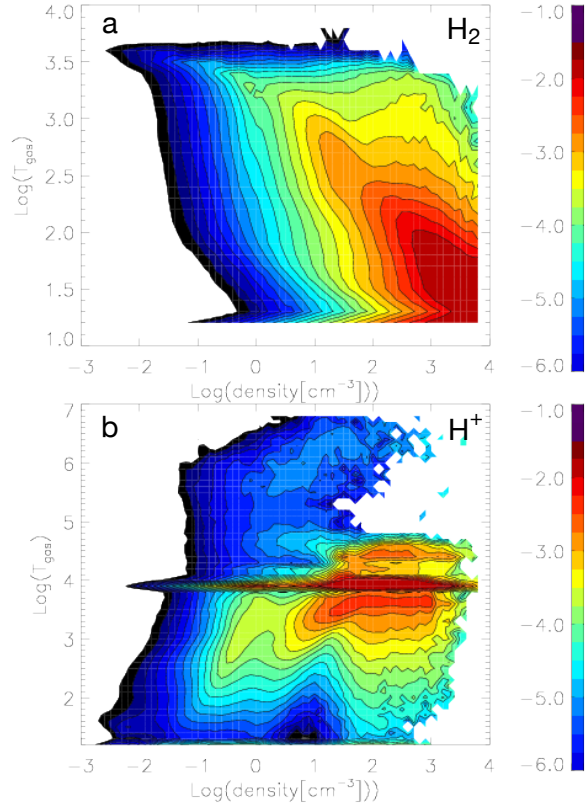}
\caption{(a) Phase-diagram for molecular hydrogen and (b) for ionized hydrogen. The contours represent mass-weighted abundances 
of H$_2$ or H$^+$ relative to their maximum values.}
\label{wada_fig: HI-HII}
\end{figure}


 \begin{figure}[h]
\centering
\includegraphics[width = 7cm]{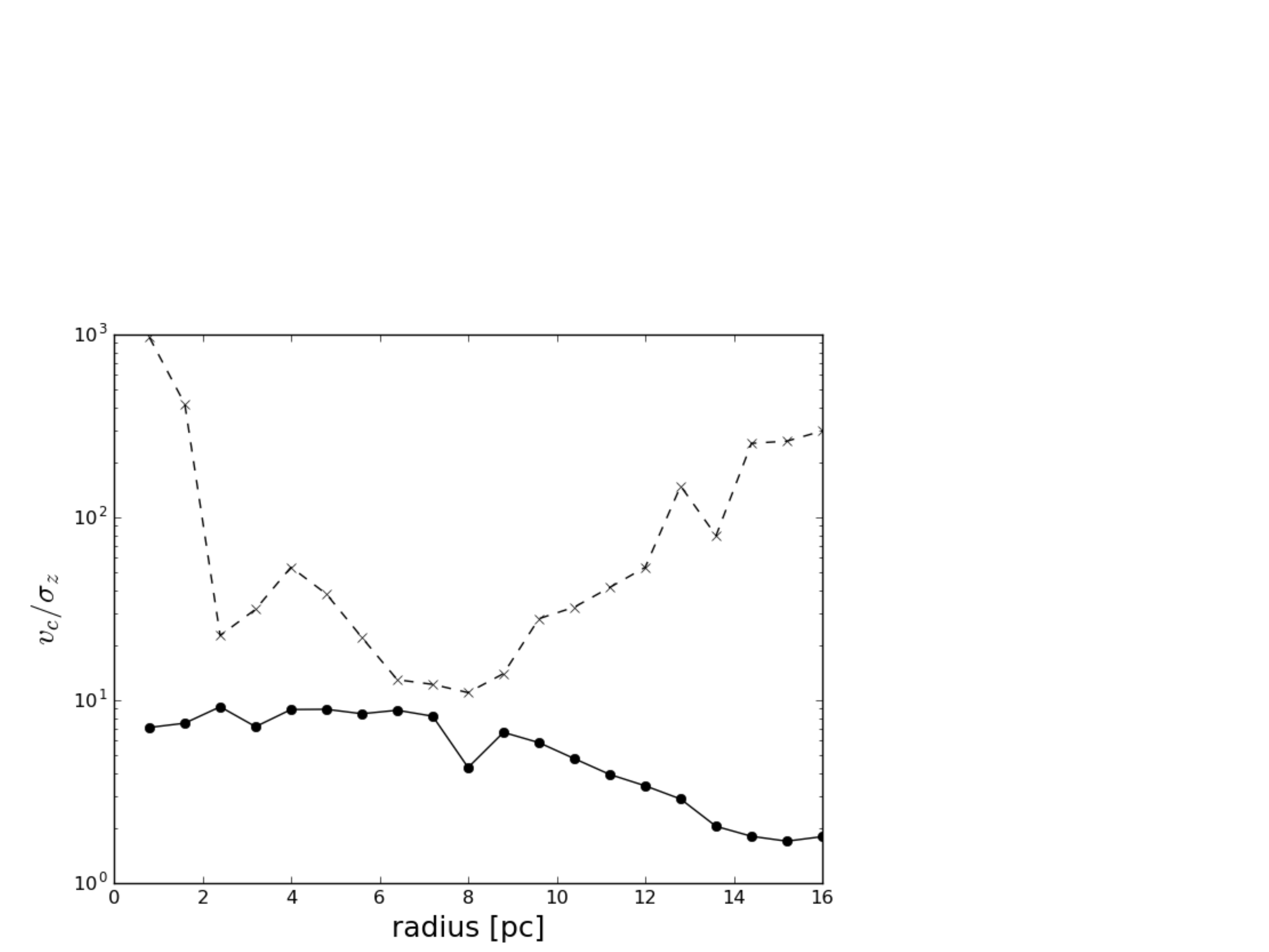} 
\caption{Radial distribution of $v_c/\sigma_z$, where $\sigma_z$ is the velocity dispersion for the $z$-direction and
$v_c$ is the rotational velocity weighted
to  the hot ($> 1000$ K) molecular hydrogen density: with SNe feedback (thick line) and without feedback (dashed line). }
\label{wada_fig: dispersion}
\end{figure}

Figure 2 shows the mass-weighted histogram of H$_2$ and H$^+$ as a function of their number density and temperature.
We find that H$_2$ mostly exists in the low temperature ($T_{gas} < 100$ K) and
high density ($n > 100$ cm$^{-3}$) phase. 
On the other hand, H$^+$ is dominant  in the hot gas around $T_{gas} \sim 8000$ K.
There are two additional  phases in the $H^+$ gas around $T_{gas}\sim 30000$ K and 5000--7000 K with $n \sim 10^2$ - $10^3$ cm$^{-3}$. These phases correspond to 
high density gases heated by the central radiation near the nucleus ($ r \lesssim 5$ pc) and 
the disk gas heated by the SNe  feedback, respectively.
In Figure 2a, hot H$_2$ ($T_{gas} \sim 1000$ -- 3000 K) is observed around 
$n \sim 10-10^3$ cm$^{-3}$.   The total H$_2$ mass in this phase is $4.7\times 10^2 M_\odot$.
 
The ratio between the rotational velocity ($v_c$) and 
the vertical velocity dispersion ($\sigma_z$) of hot ($ > 1000$ K) H$_2$ is plotted
as a function of the radius in Fig. 3. 
The ratio  declines toward  $v_c/\sigma_z \sim 1 $ at larger radii, reflecting 
the large scale height of the disk (Fig. 1) is supported by 
energy input due to SNe.  
On the other hand,  the ratio remains large ($\gtrsim10$) for the pure radiation-fountain model without SNe (dashed line).
{{ In the inner region ($r < 2$ pc),  
the vertical velocity dispersion is negligibly small compared to $v_c$ for the model without SN feedback, 
suggesting the radiation-driven fountain alone does not work to enhance the disk thickness for this low luminosity AGN (see also W15).
The energy feedback from SNe allows to keep the hot H$_2$ disk at a thickness of $\sim r/10$ even at $ r < 8 $ pc.}

{Using Very Large Telescope (VLT)/SINFONI and Keck/OSIRIS , \citet{hicks09}  found that
most local AGNs are associated with a geometrically thick, hot H$_2$ disk in the central several tens of parsecs.
They found that the molecular gas is spatially mixed with the nuclear stellar population,
and  $v_c/\sigma_z \simeq 1$ for the molecular gas, suggesting a geometrically thick disk that could obscure the nucleus.
{{ These observations are consistent with the structure in the outer part of our model with the SN feedback.}}

\begin{figure}[h]
\centering
\includegraphics[width = 12cm]{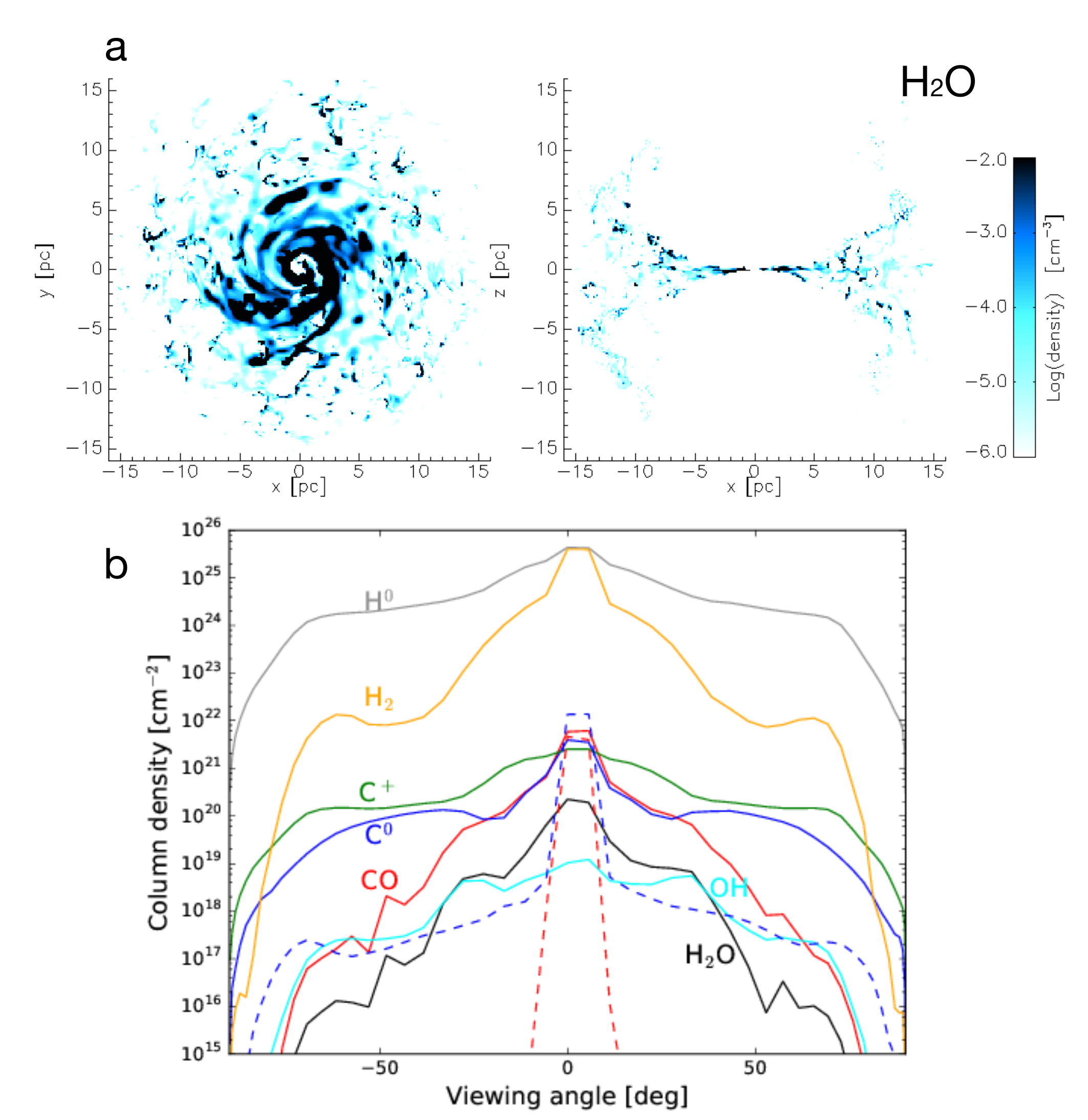} 
\caption{(a) Same as Fig. 1 but for the number density distribution of H$_2$O.  (b) Column density distribution 
as a function of the viewing angle (0$^\circ$ means edge-on) for H$^0$, H$_2$, CO, C$^0$, C$^+$, OH, and H$_2$O.  The dashed lines are
the models without supernova feedback for CO (red) and C$^0$ (blue). \vspace{1cm}}
\label{wada_fig: co-cI-h2o}
\end{figure}

%

We found that CO and C$^0$ are distributed similarly to H$_2$ and H$^0$, respectively, as shown in Fig. 1.
On the other hand, H$_2$O mostly concentrates in a thin disk at $r \lesssim 5$ pc (Fig. 4a).
This is also the case for the OH molecule.
The column density distributions for various species are
plotted as a function of the viewing angle in Fig. 4b. 
We found that the opening angles of the ionized and atomic gas disk are 
small for a given column density; in other words, their scale heights are large. 
On the other hand, molecular gases are more concentrated in the disk.
If the supernova feedback is not included, then the molecular gases form a thin, dense disk  (dashed lines).

\section{COMPARISON WITH THE NEAREST SEYFERT GALAXY}

The recent development  of new observational methods enables us to illuminate the internal structures of 
nearby AGNs, especially by using infrared and sub-millimeter/millimeter interferometers
such as VLTI, Keck, and the Atacama Large Millimeter/submillimeter Array (ALMA). 
The nearest well-observed prototypical Seyfert 2 galaxy is the Circinus galaxy.
Its black hole mass suggested by H$_2$O kinematics is $1.7\pm 0.3 \times 10^6 M_\odot$ \citep{greenhill2003}, which is 
a few orders of magnitude smaller than those in typical Seyfert galaxies, 
but close to the present  model ($M_{BH} = 2 \times 10^6 M_\odot$). 
In this section, we discuss the implications from a comparison between 
the multi-wavelength observations of the Circinus galaxy and the present model.
All numerical results below are for the model with supernova feedback, 
if it is not explicitly stated otherwise.

\subsection{Dust Emission}

Using the snapshot of the radiation-hydrodynamic model shown in \S 3, 
we run 3D Monte Carlo radiative transfer simulations for dust emission\footnote{The simulation code is RADMC-3D, http://www.ita.uni-heidelberg.de/dullemond/software/radmc-3d/}, as performed in Sch14.
Fig. 5a shows SEDs for different viewing angles.
As shown in Sch14, the SEDs are sensitive to 
the choice of the viewing angle. 
In the present case, the 10-$\mu {\rm m}$ feature appears in the emission when the viewing angles $\theta_v$ are close to
face-on ($\theta_v < 30^\circ$), but the feature appears in the absorption for $\theta_v \ge 60^\circ$.
The model SED fits the Circinus galaxy \citep{prieto2010} fairly well if $ \theta_v \gtrsim 70^\circ$ in a range between
$\lambda \sim 2 \, 	\mu\, {\rm m}$ and $\lambda \sim 60 \, \mu\, {\rm m}$.
 \citet{tristram2014} suggested that the nuclear disk-like dust emission revealed by VLT/MIDI
 has an inclination of $i \sim 75^\circ$, which is consistent with our SED analysis.
 \footnote{\citet{ruiz2001} concluded that the radius of the dusty torus in the Circinus galaxy is 2--12 pc, which is 
 consistent with our result.}

Figure 5b shows 12-$\mu {\rm m}$ images observing the system from two directions,
 i.e., 75$^\circ$ and edge-on.
Notably, the cold molecular disk observed in Fig. 1 is visible as a ``clumpy dark lane''
 in these infrared images, which obscures  the nuclear bright IR core.
 Without supernova feedback the scale height of the dense disk is not large enough to
 obscure the core\footnote{The assumed supernova rate, i.e. 0.014 yr$^{-1}$ is about a factor of ten larger than
 that suggested by the observed star formation rate in the Circinus galaxy \citep{hicks09}. However, the supernova feedback in numerical simulations with a sub-pc resolution could be less efficient than in reality \citep{gentry2016}.}.
 
We also found  that the dust emission forms a bright region that is elongated along the 
rotational axis with a dimmer counterpart in the $z < 0$ direction. 
The structures are inhomogeneous and originate from the hot dust in 
the bipolar outflows.
Recent MIR  interferometric observations of nearby Seyferts showed that 
the bulk of the MIR emission is emitted by dust in the polar region and not by the torus \citep{hoenig2013, tristram2014, asmus2016, lopez2016}. 
Our results naturally expect  the observed polar emission of dust.  The hollow cones are not actually ``empty''.
{{ However,  in the Circinus galaxy, \citet{tristram2014} also detected a disk-like MIR emission.
This might correspond to an inner part of the thin, dense disk seen in Fig. 1 at $r <  5$ pc, 
however, additional physics and  higher spatial resolution would be needed in order to study
the 3D structures of this component in our simulations. 
Note that the size of the extended polar dust emission in \citet{tristram2014} is smaller (1-2 pc) than 
the polar emission in Fig. 5b. The MIDI observation could correspond to the innermost bright 12 $\mu m$ emission in our model. 
A more detailed comparison with the MIDI data will be discussed in a subsequent paper. }

\begin{figure}[h]
\centering
\includegraphics[width = 14cm]{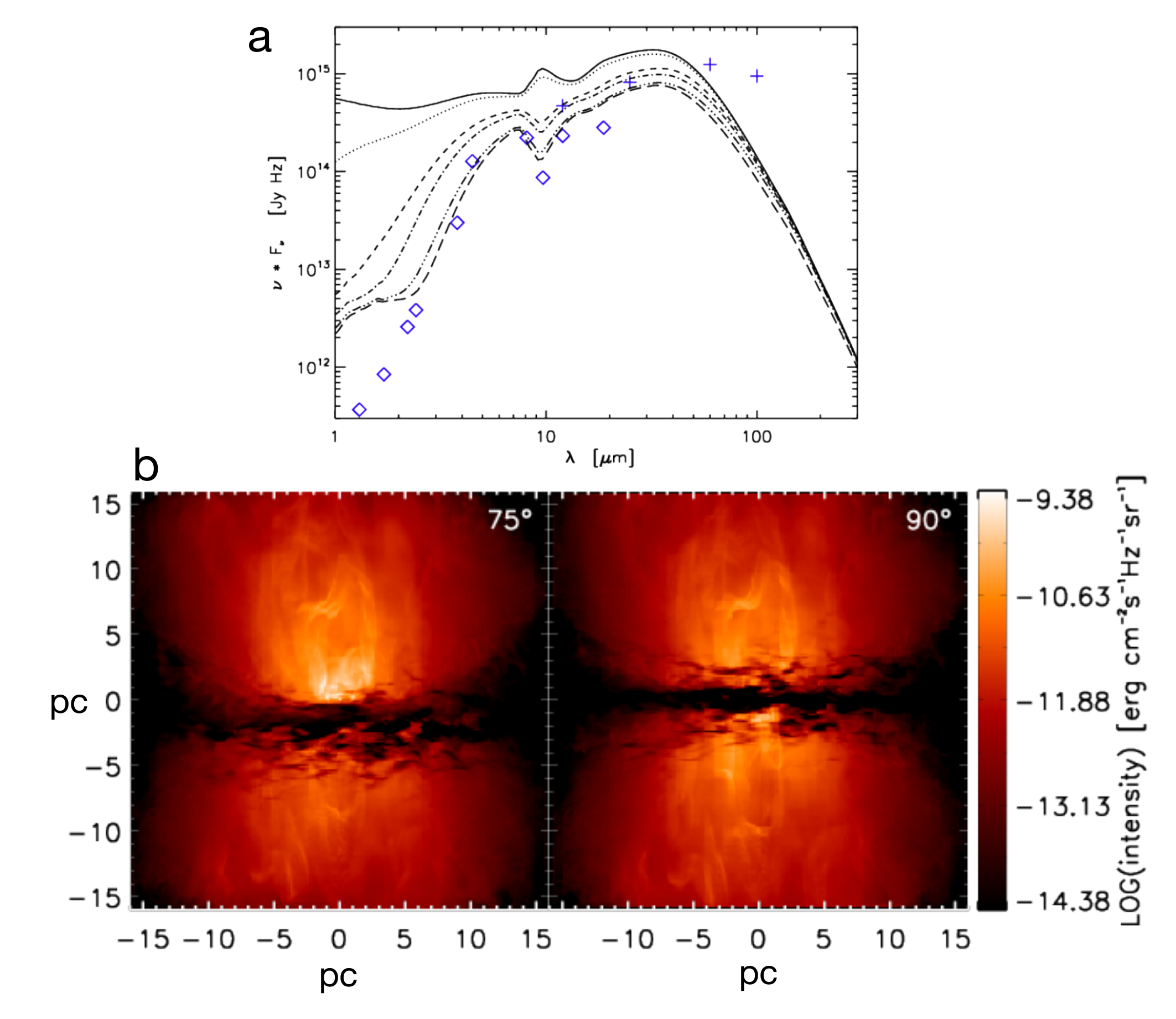}

\caption{(a)  Dust continuum radiative transfer SEDs for various inclination angles (top to bottom: 0$^\circ$, 30$^\circ$, 60$^\circ$, 70$^\circ$, 80$^\circ$, 90$^\circ$) assuming a distance to the Circinus galaxy of 4.2 Mpc. 
  The open diamonds refer to the observed nuclear
  SED, and the crosses represent large aperture IRAS data (compiled by \citealp{prieto2010}). (b) Simulated 12-$\mu\, {\rm m}$ images for an inclination angle of 75$^\circ$ (left) and for the edge-on view (right).
Labels are given in parsecs.
}
\label{wada_fig: 12um}
\end{figure}

\subsection{Cold Molecular Gas}


\citet{zhang2014} reported  
  molecular gas in the nuclear region ($r < 360$ pc) with a kinetic temperature of $\sim 200$ K and density of $\sim 10^{3.2}$ cm$^{-3}$
  using the multi-$J$ transition of $^{12}$CO observed by the Atacama Pathfinder EXperiment (APEX) telescope. 
Figure 2 shows that   there is a dominant phase of dense molecular gas, whose temperature correlates with 
density as  $ T_{gas} \sim 100$ K $(n/10^3 \; {\rm cm}^{-3})^{-1/2} $.
For $T_{gas} = 200$ K,  gas density is $n \sim 2.5 \times 10^2 {\rm cm}^{-3}$.
 As shown in Fig.1b, this high density molecular gas
forms  a clumpy disk that is affected by energy feedback from the  nuclear starburst.
Origin of the warm molecular gas is not clear in the APEX result, but 
it can be verified using high-resolution observations with ALMA in the near future.
 
 
 
 
\subsection{Hot Molecular Gas}
 
The hot molecular gas (1000--3000 K) can be traced by the near infrared region, such as  H$_2$ at 2.12 $\mu$m.
Using VLT /NaCo, \citet{mezcua2015} obtained 2.12-$\mu$m images with a resolution of 0.09--0.16 arcsec in the central 1 kpc of seven nearby Seyfert galaxies. They found that the hot H$_2$  is highly concentrated toward the central 100 pc and that its morphology is often symmetrical.
 The mass fraction of hot H$_2$ to the total cold gas mass in the central several tens of parsecs of
 nearby Seyfert galaxies is a few $\times 10^{-6}$ .
 For example,  in the Circinus galaxy,  $M_{hot} = 40 M_\odot$ for $r < 26$ pc.
In our model, the total mass of hot H$_2$ is 470 $M_\odot$, and
the total cold gas ($T \leq 40$ K) is $1\times 10^6 M_\odot$; 
therefore, the mass fraction is $4.7\times 10^{-5}$ for $r < 16$ pc, a factor of ten larger than the observation.

%
\subsection{Ionized Diffuse Gas and Compton-thick Material}
It is widely believed that the conical ionized gas structure \citep[][and references therein]{prieto2005} provides indirect evidence for the existence of a geometrically and optically thick tori. 
Our model, however naturally reproduces a double-hollow cone structure of ionized gas  as a result of the radiation-driven outflows
without a torus.
The opening angle of the ionized gas is about 120$^\circ$ (Fig. 1c),
which is similar to the observed ionization cone in the Circinus galaxy \citep{ marconi1994,veilleux1997,
greenhill2003}.

We found that the ionized gas  in the biconical outflows  is not uniform with 
$n \sim 10-100$ cm$^{-3}$  (Fig. 1c). 
This is similar to the situation expected in a model of the coronal line region of Seyfert galaxies \citep[e.g.][]{murayama1998}.
The spectroscopic properties of the ionized gas in our models will be studied in detail by 
applying {\it Cloudy} \citep{ferland2013} to the hydrodynamic model in a subsequent paper.

\citet{marinucci2013} found a Compton thick, clumpy structure that 
is axisymmetric with respect to the nucleus of the Circinus galaxy, producing the cold reflection and the iron K$\alpha$ line.
They suggested that this could be the outer part of the dusty region responsible for the infrared emission. 
\citet{arevalo2014}  suggested that the ``torus" has an equatorial column density $N_{\rm H} = 6-10\times 10^{24}$ cm$^{-2}$, 
and the intrinsic X-ray luminosity is $L_{2-10 \, {\rm keV}} = 2.3-5.1 \times 10^{42}$ erg s$^{-1}$ using {\it NuSTAR, XMM-Newton},
and {\it Chandra}. The column density and X-ray luminosity are consistent with our model.
Our model shows that the Compton thick gas ($N_{\rm H} > 1.5 \times 10^{24}$ cm$^{-2}$) obscures the nucleus, if the viewing angle is
$\sim 75^\circ$. 
Therefore, the cold reflection gas found in the X-ray observations would be the clumpy gas in the outer
part of the circumnuclear disk shown in the H$_2$ map (Fig. 1b).

\acknowledgments

The authors are grateful to the anonymous referee for his/her constructive comments and suggestions. We thank Dr. Takuma Izumi for his valuable comments.
Numerical computations were performed on a Cray XC30 at the Center for Computational Astrophysics at the National Astronomical Observatory of Japan. We thank C. P. Dullemond for making RADMC-3D publicly
available and for his continuous support with the code.
KW was supported by JSPS KAKENHI Grant Number 16H03959.

\end{document}